%% file: colm2024_conference.tex
\documentclass{article} 
\usepackage{colm2024_conference}

\usepackage{microtype}
\usepackage{hyperref}
\usepackage{url}
\usepackage{booktabs}
\usepackage{graphicx}
\usepackage{setspace}
\usepackage{enumitem}
\usepackage{times}
\usepackage{latexsym}
\usepackage{booktabs}  

\usepackage[T1]{fontenc}

\usepackage[utf8]{inputenc}

\usepackage{microtype}

\usepackage{inconsolata}
\usepackage{fancyhdr}

\usepackage{graphicx}
\usepackage{subcaption}
\usepackage{multirow}  
\usepackage{tcolorbox}
\usepackage{array} 
\usepackage{amsmath} 
\usepackage{microtype}

\title{CodecBench: A Comprehensive Benchmark for Acoustic and Semantic Evaluation}

\author{
Ruifan Deng$^{1}$ \hspace{.3em}
Yitian Gong$^{1,2}$ \hspace{.3em}
Qinghui Gao$^{1,2}$ \hspace{.3em}
Luozhijie Jin$^{1}$ \hspace{.3em}
\\
\textbf{
Qinyuan Cheng$^{1}$ \hspace{.3em}
Zhaoye Fei$^{1}$ \hspace{.3em}
Shimin Li$^{1}$ \hspace{.3em}
Xipeng Qiu$^{1, 2}$\thanks{ {} Corresponding author.}
}
\\
\texttt{\{rfdeng23, ytgong24\}@m.fudan.edu.cn} \quad \texttt{xpqiu@fudan.edu.cn} \\ 
[1ex]
$^{1}$Fudan University \\
$^{2}$Shanghai Innovation Institute \\
}

\colmfinalcopy

\fancypagestyle{firstpage}{
  \lhead{\begin{picture}(0,0)\put(0,-6){\includegraphics[width=0.3\linewidth]{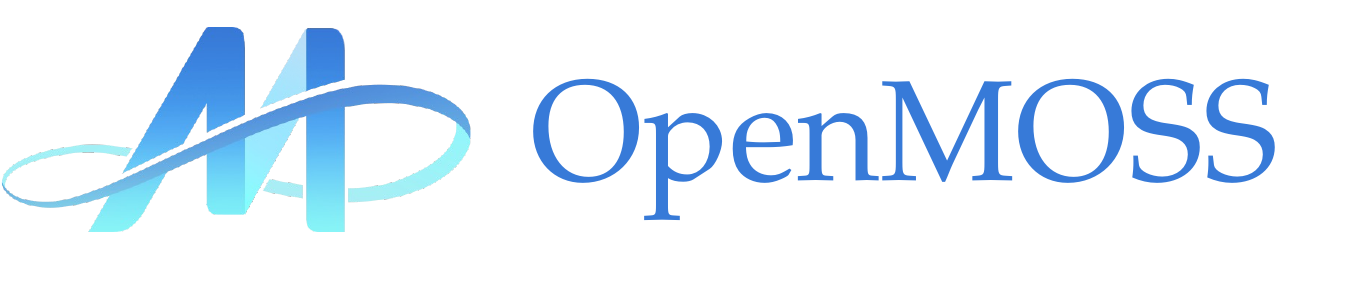}}\end{picture}}
}

\begin{document}

\maketitle

\thispagestyle{firstpage}

\begin{abstract}
\input{000_Abstract}
\end{abstract}

\section{Introduction}
\label{intro}
\input{100_Introduction}

\section{Related Work}
\label{related work}
\input{200_Related_Work}

\section{CodecBench}
\label{codecbench}
\input{300_CodecBench}

\section{Experiments}
\label{experiments}
\input{400_Experiments}

\section{Limitations}
\label{limits}
\input{500_Limitaions}

\section{Conclusion}
\label{conclusion}
\input{600_Conclusion}

\bibliography{colm2024_conference}
\bibliographystyle{colm2024_conference}

\newpage
\appendix
\section{Appendix}
\input{900_Appendix}

\end{document}

%% file: 000_Abstract.tex
With the rise of multimodal large language models (LLMs), audio codec plays an increasingly vital role in encoding audio into discrete tokens, enabling integration of audio into text-based LLMs. Current audio codec captures two types of information: acoustic and semantic. As audio codec is applied to diverse scenarios in speech language model , it needs to model increasingly complex information and adapt to varied contexts, such as scenarios with multiple speakers, background noise, or richer paralinguistic information. However, existing codec's own evaluation has been limited by simplistic metrics and scenarios, and existing benchmarks for audio codec are not designed for complex application scenarios, which limits the assessment performance on complex datasets for acoustic and semantic capabilities. We introduce CodecBench, a comprehensive evaluation dataset to assess audio codec performance from both acoustic and semantic perspectives across four data domains. Through this benchmark, we aim to identify current limitations, highlight future research directions, and foster advances in the development of audio codec. The codes are available at \url{https://github.com/RayYuki/CodecBench}.

%% file: 100_Introduction.tex
\begin{figure}[ht]
  \centering
  \includegraphics[width=1.0\linewidth]{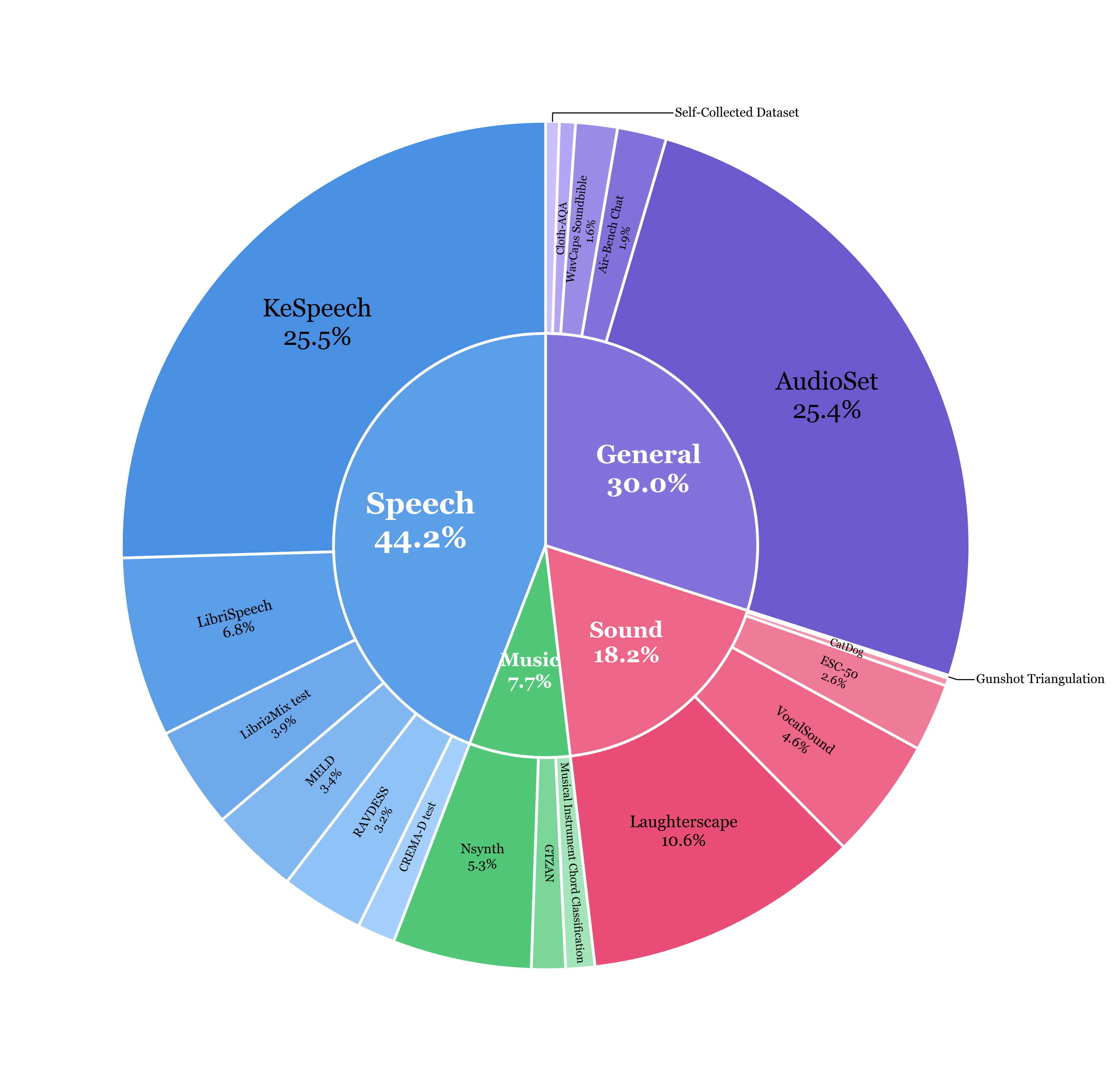}
  \caption{CodecBench data distribution overview, containing 18 open-source datasets and 1 self-collected dataset, including 6 speech datasets, 3 music datasets, 5 sound datasets and 5 general audio datasets. } 
  \label{fig:codecbench_data_distribution} 
\end{figure}

Recent advancements in text-based large language models (LLMs), such as ~\citep{sun2024moss, yang2024qwen2}, have promoted the rapid development of speech language models ~\citep{defossez2024moshi, li2025baichuan}, of which audio codec plays a role of no less importance than the text-based LLMs. By encoding audio signals into discrete tokens, the audio codec bridges the gap between audio and text, allowing text-based LLMs to process audio input as text. However, unlike text, audio contains more information than semantics, including emotion, speaker identity, and general audio, proposing unique challenges for modeling and evaluation.

Audio codec can capture both acoustic information and semantic information, which is crucial for its integration into speech language models. This representation supports a wide range of applications, from speech synthesis to cross-modal reasoning. As speech language models are deployed in increasingly complex scenarios, such as noisy environments, multilingual settings, or emotionally interactions, audio codec must model diverse and complex audio inputs. However, comprehensively evaluating the performance of the codec in these dimensions including semantic evaluation across multiple domains, such as speech and sound, remains an open challenge.

Despite the strong reconstruction performance of existing audio codecs ~\citep{kumar2023high, zhang2024speechtokenizerunifiedspeechtokenizer, defossez2024moshi}, their evaluation has been limited by simplistic metrics and benchmarks. Existing codec benchmarks, often relying on datasets like LibriSpeech~\citep{panayotov2015librispeech}, typically lack background noise and expressive variability, creating a gap between controlled settings and real-world demands. For instance, while Codec-SUPERB~\citep{wu2024codec} incorporates datasets across multiple domains, these are predominantly clean, limiting its ability to assess audio codec performance in practical scenarios. Moreover, some benchmarks fail to adequately evaluate semantic aspects critical for speech language models.

To address these challenges, we introduce CodecBench, a comprehensive evaluation framework designed to assess audio codec performance from both acoustic and semantic perspectives across diverse scenarios. The main features of CodecBench are as follows:
\begin{itemize}
\item CodecBench provides a rich evaluation dataset comprising numerous open-source and self-collected datasets, covering a wide range of audio scenarios, including multilingual speech, noisy environments, and emotionally expressive audio, satisfying the requirements of speech language models; 

\item CodecBench employs a diverse set of evaluation metrics and methods encompassing acoustic and semantic dimensions, incorporating  more acoustic metrics, a embedding-based classifier method inspired by ARCH, and an ASR task adapted from the SUPERB framework to assess semantic information, extending beyond previous benchmarks;

\item CodecBench conducts an in-depth evaluation of existing codecs, identifying their limitations and proposing potential directions for improvement. 
\end{itemize}
Through this benchmark, we aim to bridge the evaluation gap and foster advancements in audio codec development for multimodal applications.

%% file: 200_Related_Work.tex
\subsection{Speech Language Model}

In recent years, LLMs have demonstrated strong natural language processing abilities. With the integration of the audio codec, speech language models have reached rapid advancements. Models such as AudioLM~\citep{borsos2023audiolm} employ autoregressive (AR) transformers and hierarchical modeling techniques to directly process and learn from audio data. Similarly, VALL-E~\citep{wang2023neural} and SPEAR-TTS~\citep{kharitonov2023speak} leverage AR transformers to frame text-to-speech tasks as audio modeling problems. SpeechGPT~\citep{zhang2023speechgpt}, Moshi~\citep{defossez2024moshi}, and GLM4-Voice~\citep{zeng2024glm} explore speech-to-speech dialogue tasks. In these models, the audio codec holds a pivotal position in enabling efficient audio processing and representation.

\subsection{Audio Codec}

Typically trained for acoustic-level reconstruction tasks, an audio codec comprises an encoder, a quantizer, and a decoder. The encoder compresses input audio into latent features, the quantizer converts these into discrete representations, and the decoder reconstructs the audio. This architecture enables efficient representation learning on large-scale datasets, supporting unified processing of diverse audio types with robust reconstruction performance. Recent advancements have focused on three key areas: (1) better reconstruction quality, like DAC~\citep{kumar2023high}, MaskGCT Codec ~\citep{wang2024maskgct}; (2) lower bitrate, such as BigCodec~\citep{xin2024bigcodec} and Stable-Codec~\citep{parker2024scaling}; and (3) richer semantic information, such as  SpeechTokenizer~\citep{zhang2024speechtokenizerunifiedspeechtokenizer}, X-Codec-2.0~\citep{ye2025llasascalingtraintimeinferencetime}, and Mimi~\citep{defossez2024moshi}.  A high-performing codec must balance these dimensions to meet the demands of the real world.

\subsection{Evaluation Benchmarks}

As audio codec technology has advanced, several benchmarks have been developed to evaluate its performance. For instance, Codec-SUPERB~\citep{wu2024codec} assesses both signal-level reconstruction and downstream task performance. ESPnet-Codec~\citep{shi2024espnet} integrates multiple codecs into a unified training and evaluation framework, VERSA~\citep{shi2024versa}, extending evaluation to generative tasks such as text-to-speech (TTS) and singing voice synthesis (SVS). Closest to our work is Codec-SUPERB; however, it has notable limitations. For acoustic evaluation, it employs a limited set of metrics and relies on relatively clean datasets that fail to address real-world application demands. For semantic evaluation, it uses various pretrained models to directly measure reconstructed audio performance on downstream tasks. However, with the rapid development of speech language models, this approach fails to evaluate the semantic information that the audio codec conveys to the language model, including alignment with text and paralinguistic information. These shortcomings underscore the need for a comprehensive benchmark that evaluates both acoustic and semantic performance across diverse real-world scenarios.

%% file: 300_CodecBench.tex
To comprehensively evaluate the performance of the audio codec, we first utilize codec models to resynthesize the datasets described in Section~\ref{codecbench:datasets}. The acoustic metrics described in Section~\ref{codecbench:acoustic metric} are used to assess the quality of the codec reconstruction. Furthermore, we utilize the codec embedding output, as detailed in Section~\ref{codecbench:semantic metrics}, to evaluate its semantic performance.

\subsection{Datasets}
\label{codecbench:datasets}
Many audio codec models have been evaluated under their respective experimental settings, yet most fail to account for comprehensive real-world usage scenarios, relying on datasets with limited coverage. Therefore, we select multiple datasets across three primary audio categories: speech, sound, and music, and include comprehensive general audio datasets that combine these categories, ensuring alignment with real-world application needs. The distribution of datasets is illustrated in Figure~\ref{fig:codecbench_data_distribution}, detailed information for datasets is provided in Appendix \ref{appendix:datasets_details}.

\textbf{Speech}: Speech refers to human-generated audio that conveys linguistic content through a language system, utilizing phonemes and prosody to express semantic meaning. As the most common scenario for audio codec applications, we selected diverse speech datasets encompassing linguistic diversity, speaker diversity, emotional diversity, and complex scenarios (e.g., multi-speaker dialogues). These datasets enable robust evaluation of the ability of audio codec to encode semantic and acoustic properties for speech language model tasks such as automatic speech recognition (ASR) and text-to-speech (TTS).

\textbf{Music}: Music refers to audio structured by melody, rhythm, or harmony, typically produced by instruments, vocals, or electronic synthesis. Note that purely human vocalizations with melody are categorized under speech datasets. Our music datasets span diverse genres, from classical to contemporary, including instrumental compositions, pure music tracks, and songs with vocals.

\textbf{Sound}: Sound refers to non-linguistic, non-musical audio produced by humans, animals, or environmental sources, typically lacking explicit linguistic structure but sometimes can convey contextual or emotional information. We collected diverse sound datasets covering human non-verbal vocalizations, animal sounds, and environmental sounds. These datasets are essential for evaluating the performance of audio codec in noisy environments and emotionally nuanced interactions, enhancing its robustness in real-world scenarios.

\textbf{General Audio}: General audio is a broad category that encompasses any audio signal, including combinations of speech, music, and sound, reflecting complex real-world scenarios. To address practical demands, we selected datasets comprising multilingual speech, varied contexts, and mixed audio sources (e.g., audios with dialogue, music, and ambient noise). In addition to these selected open-source datasets, we introduce a self-collected general audio dataset featuring more exaggerated expressiveness or more complex speaker scenarios, posing higher performance demands on audio codecs. These datasets assess the ability of audio codec to extract and reconstruct information in diverse scenarios, ensuring to meet the growing needs of speech language models.

\subsection{Metrics}

To comprehensively assess the audio codec’s performance, we employ a diverse set of metrics covering both acoustic and semantic dimensions, tailored to the requirements of speech language models and real-world applications.

\subsubsection{Acoustic Metrics}
\label{codecbench:acoustic metric}
Acoustic metrics evaluate the fidelity and perceptual quality of reconstructed audio, focusing on signal-level accuracy and human perception. The following metrics, summarized in Table~\ref{tab:acoustic_metrics}, are used, with arrows indicating whether higher ($\uparrow$) or lower ($\downarrow$) values are desirable.

\input{Acoustic_Metrics}

\subsubsection{Semantic Metrics}
\label{codecbench:semantic metrics}
Semantic metrics assess the preservation of linguistic and contextual information, crucial for the integration of the audio codec into speech language models. We employ the following metrics:

\label{asr_probing_task}
\textbf{ASR Probing Task} To evaluate the semantic alignment between the codec and text, we employ an ASR probing task, adapted from the SUPERB framework~\citep{yang2021superb}, to assess the semantic quality of tokenized representations. We trained a downstream ASR model using quantized embeddings, with the pretrained codec fixed. These quantized embeddings are upsampled to a minimum frame rate of 50 Hz via replication before being fed into a downstream model (see Appendix~\ref{appendix:asr_probing_details} for details). The downstream model comprises a two-layer bidirectional LSTM optimized with CTC loss for character-level prediction~\citep{hochreiter1997long,graves2006connectionist}. All models are trained on the LibriSpeech train-clean-100 subset and evaluated on the LibriSpeech dev-clean subset~\citep{panayotov2015librispeech}, using Word Error Rate (WER) as the metric for semantic performance, where a lower WER indicates better semantic alignment. The ASR probing task experiments are conducted with a batch size of 4, a maximum learning rate of $1\times10^{-4}$, and training for 400,000 steps.

\textbf{Classification Task}: The ASR Probing Task primarily evaluates textual semantic information, missing other contextual and emotional aspects. Inspired by ARCH~\citep{la2024benchmarking}, we extract embeddings from the audio codec and train classifiers on labeled datasets (selected from Section~\ref{codecbench:datasets}) to assess the codec’s ability to preserve contextual and emotional information, reflecting its semantic expressiveness beyond textual content. The accuracy of these classifiers on the test split of the datasets serves as the evaluation metric. This method also applies to evaluating tokenizers from ASR and SSL models. Details on the dataset categories, label counts, and other information are presented in Table~\ref{tab:semantic_datasets}. The classification task experiments are conducted with a batch size of 16, a maximum learning rate of $1\times10^{-3}$, and training for 20 epochs for each dataset.

\input{Semantic_Datasets}

%% file: Acoustic_Metrics.tex
\begin{table*}[ht]
\centering
\hyphenation{Per-cep-tu-al Short-Time Vir-tu-al}
\footnotesize
\setlength{\tabcolsep}{3pt}
\renewcommand{\arraystretch}{0.5}
\begin{tabular}{p{3.5cm} p{4.5cm} p{2.0cm} c}
\vspace{0.00005pt} \\[-0.00005pt]
\toprule
\textbf{Metric} & \textbf{Description} & \textbf{Range} & \textbf{Ref.} \\
\midrule
Mel Loss & Measures Mel spectrogram difference for perceptual quality. & [0, $\infty$] (↓) & \\
\vspace{0.00005pt} \\[-0.00005pt]
\hline
\vspace{0.00005pt} \\[-0.00005pt]
Speech Quality (PESQ) & Measures speech quality for perceptual accuracy. & [1, 4.5] (↑) & ~\citep{rix2001perceptual}\\
\vspace{0.00005pt} \\[-0.00005pt]
\hline
\vspace{0.00005pt} \\[-0.00005pt]
Spectral Convergence (SC) & Measures spectral feature difference for frequency accuracy. & [0, $\infty$] (↓) & ~\citep{sturmel2011signal} \\
\vspace{0.00005pt} \\[-0.00005pt]
\hline
\vspace{0.00005pt} \\[-0.00005pt]
Signal-to-Distortion Ratio (SDR) & Measures signal-to-distortion ratio for audio quality. & [$-\infty$, $\infty$] (↑) & ~\citep{fevotte2005bss_eval} \\
\vspace{0.00005pt} \\[-0.00005pt]
\hline
\vspace{0.00005pt} \\[-0.00005pt]
Scale-Invariant SDR (SI-SDR) & Measures scale-invariant signal quality for robustness. & [$-\infty$, $\infty$] (↑) & ~\citep{le2019sdr} \\
\vspace{0.00005pt} \\[-0.00005pt]
\hline
\vspace{0.00005pt} \\[-0.00005pt]
Speaker Similarity (SIM) & Measures speaker identity retention for speech synthesis. & [0, 1] (↑) & ~\citep{toda2016voice} \\
\vspace{0.00005pt} \\[-0.00005pt]
\hline
\vspace{0.00005pt} \\[-0.00005pt]
Short-Time Objective Intelligibility (STOI) & Measures speech intelligibility for noisy environments. & [0, 1] (↑) & ~\citep{taal2010short} \\
\vspace{0.00005pt} \\[-0.00005pt]
\hline
\vspace{0.00005pt} \\[-0.00005pt]
Virtual Speech Quality Objective Listener (ViSQOL) & Measures speech quality for human perception. & [1, 5] (↑) & ~\citep{chinen2020visqol} \\
\bottomrule
\end{tabular}
\caption{Acoustic Metrics for CodecBench.}
\label{tab:acoustic_metrics}
\end{table*}

%% file: Semantic_Datasets.tex
\begin{table*}[ht]
\centering
\hyphenation{recog-ni-tion}
\footnotesize
\setlength{\tabcolsep}{2pt}
\renewcommand{\arraystretch}{0.5}
\begin{tabular}{p{2cm} p{5cm} c}
\vspace{0.00005pt} \\[-0.00005pt]
\toprule
\textbf{Category} & \textbf{Dataset} & \textbf{Number of Labels} \\
\midrule
       & RAVDESS & 8 \\[3pt]
Speech & CREMA-D & 6 \\[3pt]
       & MELD & 7 \\
\vspace{0.00005pt} \\[-0.00005pt]
\hline
\vspace{0.00005pt} \\[-0.00005pt]
      & GTZAN & 10 \\[3pt]
Music & Musical Instrument Chord Classification & 2 \\[3pt]
      & Nsynth & 10 \\
\vspace{0.00005pt} \\[-0.00005pt]
\hline
\vspace{0.00005pt} \\[-0.00005pt]
Sound & ESC-50 & 50 \\[3pt]
      & VocalSound & 6 \\
\bottomrule
\end{tabular}
\caption{Datasets for Semantic Evaluation in CodecBench.}
\label{tab:semantic_datasets}
\end{table*}

%% file: 400_Experiments.tex
\subsection{Experimental Setup}

We selected multiple open-source audio codec models, including Mimi of Moshi~\citep{defossez2024moshi}, DAC~\citep{kumar2023high}, MaskGCT~\citep{wang2024maskgct}, FlowDec~\citep{welker2025flowdec}, BigCodec~\citep{xin2024bigcodec}, X-Codec-2.0 of LLaSA~\citep{ye2025llasascalingtraintimeinferencetime}, Stable-Codec~\citep{parker2024scaling}, and Baichuan-Audio's tokenizer~\citep{li2025baichuan}. These models were chosen to cover a diverse range of architectures and training methods, resulting in a total of 14 audio codec models for acoustic evaluation. A brief overview of these models is provided in Table~\ref{tab:codecbench_performance}. SIM is calculated as the cosine similarity between speaker embeddings extracted from original and reconstructed audio using a pre-trained speaker verification model\footnote{\url{https://github.com/microsoft/UniSpeech/tree/main/downstreams/speaker_verification}}. For semantic evaluation, considering the need to use embeddings from model outputs, we tested a subset of the aforementioned models, using a single NVIDIA H100 GPU for each codec model.

\input{Acoustic_Evalution}

\subsection{Acoustic Evaluation}

Table~\ref{tab:codecbench_performance} presents the performance of various audio codecs in datasets within the CodecBench framework.To ensure fair evaluation, all audio was resampled to 16 kHz. The DAC-24k-rvq32 variant achieves superior performance in most metrics, including PESQ, STOI, SIM, and ViSQOL, but is outperformed by DAC-44k-rvq9 in MSE, SDR, and SI-SDR. Notably, for SDR and SI-SDR the difference may be large despite other metrics being not much different within the same model architecture like DAC. We attribute this to higher frame rates that improve temporal resolution, which reduces temporal errors and increases the signal-to-distortion ratio, thereby improving SDR and SI-SDR. At lower bitrates, Stable-Codec variants consistently underperform compared to other codecs, while BigCodec achieves the best performance in single-codebook settings. At higher bitrates, DAC variants have a leading position. FlowDec and Baichuan-Audio Tokenizer, both leveraging flow matching, exhibit competitive performance in the same bitrate setting. 

\input{Selected_Metric}

Table~\ref{tab:selected_comprehensive} presents the performance of these codec models across four dataset categories based on Mel Loss, PESQ-NB, and STOI metrics, showing a clear comparison of model performance across different data domains. At high bitrates, the DAC series demonstrates superior performance across all domains. Notably, FlowDec exhibits slightly better overall performance than DAC-24k-rvq8, under identical codebook and bitrate configurations, particularly on the Music datasets. This highlights the advantages of the flow-matching approach. However, FlowDec’s STOI scores are consistently lower, which may be attributed to its using a self-trained DAC-structured model. At low bitrates, BigCodec achieves excellent performance on the Speech datasets but underperforms compared to Mimi-8 in other domains. Furthermore, compared to high-bitrate settings, codec models struggle more significantly to model non-vocal data, such as Music and Sound, at low bitrates, with a more pronounced performance drop relative to the Speech datasets. Detailed results are provided in Appendix~\ref{appendix:Additional experiment results}.

\subsection{Semantic Evaluation}

Given that both of the semantic evaluation methods that we adopt are based on embeddings, we only tested a subset of the models mentioned earlier. For instance, FlowDec, which enhances DAC through flow matching, exhibits embeddings that are not significantly different from DAC.

\begin{figure}[ht]
  \centering
  \includegraphics[width=0.7\linewidth]{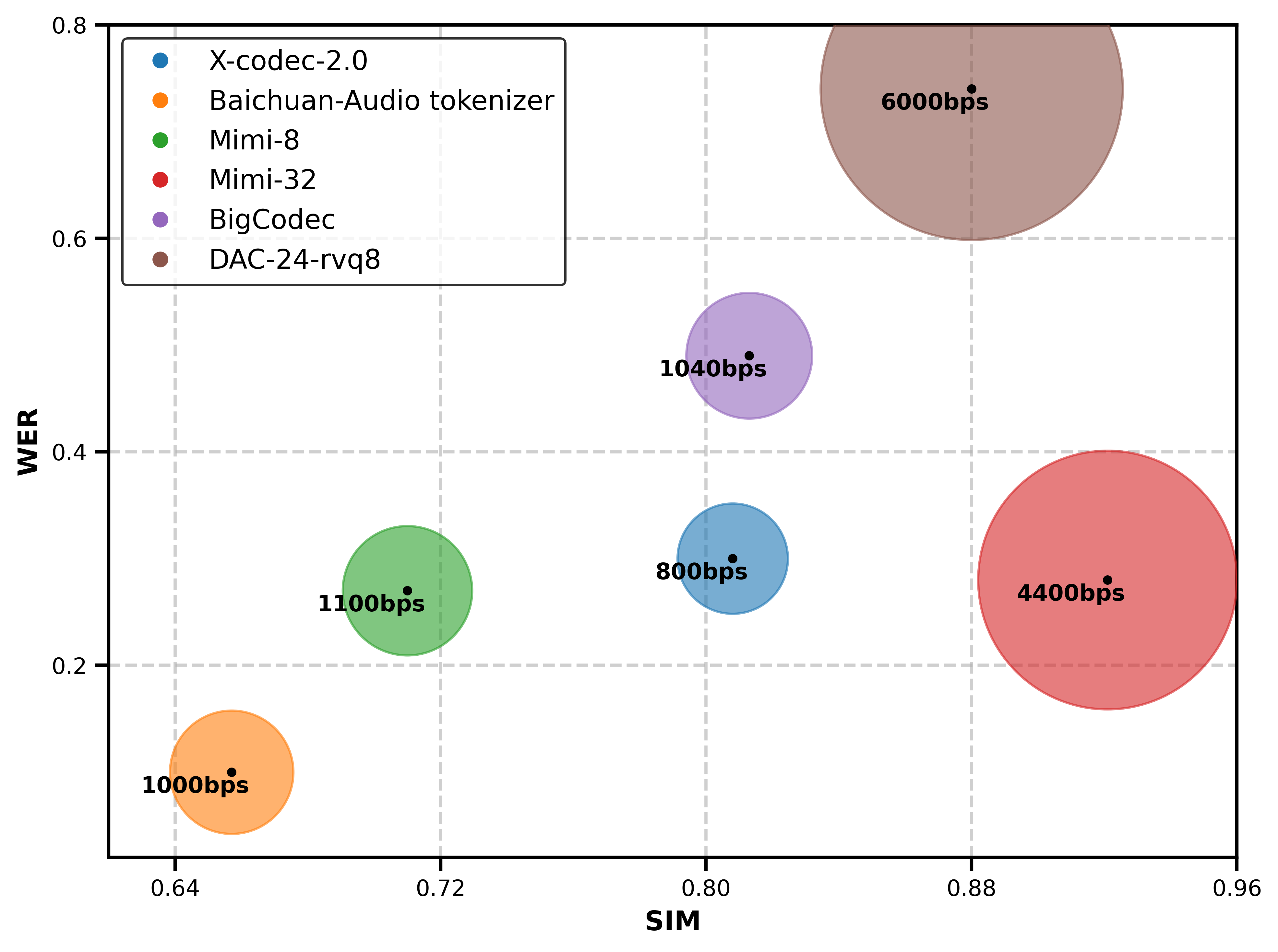} 
  \caption{Comparisons between different codecs on WER and SIM.} 
  \label{fig:wer_sim} 
\end{figure}

\subsubsection{ASR Probing Task}
Figure~\ref{fig:wer_sim} presents a comparison of Word Error Rate (WER) and Speaker Similarity (SIM) for audio codec models on the LibriSpeech dataset, plotted with the models' bitrate as the area of the circle. In particular, DAC and BigCodec, which perform well in acoustic evaluation at high and low bitrates, respectively, exhibit significantly poorer WER performance compared to other models. This can be attributed to the absence of semantic information incorporation during their pre-training phase. Among models with similar WER, X-codec-2.0, which operates with a single codebook and a lower bitrate, achieves a higher SIM than Mimi-8, showing its superior performance in this metric.

\subsubsection{Classification Task}
Table~\ref{tab:classification_performance} reports the classification accuracy results in multiple datasets that span three categories: speech, music, and sound. For comparison, we include the Whisper-small~\citep{radford2022robustspeechrecognitionlargescale} performance as a baseline. Hidden size refers to the embedding dimension of codec output used for classifier training. Despite its strong WER performance, X-codec-2.0 performs poorly on this task, even falling behind DAC. We hypothesize that X-codec-2.0's single-codebook, low-bitrate structure prioritizes preserving text-related semantic information at the expense of other semantic features. In contrast, models with larger codebooks, even without explicit training for semantic information, can still capture paralinguistic information. Compared to Whisper-small, Mimi and Baichuan-Audio tokenizer achieve better performance on the music dataset, though they remain slightly inferior on the speech and sound datasets.

\input{Classification_Eval}

\subsubsection{Token-Based Experiment}
In addition to embedding-based experiments, we explored semantic evaluation using tokens from the quantizers of audio codec models. Compared to the embedding-based approach, the token-based method requires training additional embeddings to map tokens into a unified dimensional space for subsequent training. However, we found that for models with a single large codebook, such as Stable-Codec and X-codec-2.0, training an embedding mapping from tens of thousands of dimensions to 1024 using only the LibriSpeech dataset was nearly infeasible. For other models like Mimi, which are more amenable to training, the results on both WER and SIM metrics using tokens align with the overall trends observed in the embedding-based experiments.

%% file: Acoustic_Evalution.tex
\begin{table}[ht]
  \centering
  \footnotesize
  \setlength{\tabcolsep}{3pt}
  \renewcommand{\arraystretch}{0.5}
  \resizebox{\textwidth}{!}{%
  \begin{tabular}{l c c c c c c c c c c c c c}
    \noalign{\global\arrayrulewidth 1pt}\hline\noalign{\global\arrayrulewidth 0.4pt}
    \vspace{0.05pt} \\[-0.05pt]
    \textbf{Model} & \textbf{Sample Rate} & \textbf{nq} & \textbf{kbps} & \textbf{Mel Loss$\downarrow$} & \textbf{MSE$\downarrow$} & \textbf{PESQ-NB$\uparrow$} & \textbf{PESQ-WB$\uparrow$} & \textbf{SC$\downarrow$} & \textbf{SDR$\uparrow$} & \textbf{SI-SDR$\uparrow$} & \textbf{SIM$\uparrow$} & \textbf{STOI$\uparrow$} & \textbf{ViSQOL$\uparrow$} \\
    \vspace{0.05pt} \\[-0.05pt]
    \hline
    \vspace{0.05pt} \\[-0.05pt]
    DAC & 24000 & 8 & 6.0 & 0.687 & 0.017 & 3.474 & 3.067 & 0.302 & 4.226 & 1.621 & 0.840 & 0.836 & 4.180 \\[3pt]
    DAC & 24000 & 32 & 24.0 & \textbf{0.335} & 0.014 & \textbf{4.383} & \textbf{4.367} & \textbf{0.134} & 6.143 & 1.748 & \textbf{0.962} & \textbf{0.956} & 4.373 \\[3pt]
    DAC & 44000 & 9 & 7.74 & 0.618 & \textbf{0.001} & 3.880 & 3.546 & 0.235 & \textbf{11.111} & \textbf{10.410} & 0.894 & 0.873 & \textbf{4.380} \\
    \vspace{0.05pt} \\[-0.05pt]
    \hline
    \vspace{0.05pt} \\[-0.05pt]
    MaskGCT Codec & 24000 & 8 & 4.0 & 1.056 & 0.014 & 2.634 & 2.017 & 0.547 & -2.673 & 1.633 & 0.793 & 0.763 & 4.031 \\[3pt]
    MaskGCT Codec & 24000 & 12 & 6.0 & 1.021 & 0.013 & 2.758 & 2.125 & 0.532 & -2.196 & 1.813 & 0.822 & 0.785 & 4.065 \\
    \vspace{0.05pt} \\[-0.05pt]
    \hline
    \vspace{0.05pt} \\[-0.05pt]
    BigCodec & 16000 & 1 & 1.04 & 1.351 & 0.016 & 2.123 & 1.644 & 0.630 & -4.737 & 1.076 & 0.622 & 0.656 & 3.793 \\
    \vspace{0.05pt} \\[-0.05pt]
    \hline
    \vspace{0.05pt} \\[-0.05pt]
    Mimi & 24000 & 8 & 1.1 & 1.426 & 0.007 & 2.196 & 1.700 & 0.506 & 0.730 & 2.640 & 0.670 & 0.661 & 3.801 \\[3pt]
    Mimi & 24000 & 32 & 4.4 & 1.210 & 0.004 & 2.987 & 2.478 & 0.344 & 6.339 & 6.003 & 0.834 & 0.799 & 3.992 \\
    \vspace{0.05pt} \\[-0.05pt]
    \hline
    \vspace{0.05pt} \\[-0.05pt]
    Stable-Codec & 16000 & 1 & 0.4 & 2.224 & 0.012 & 1.899 & 1.477 & 1.642 & -0.629 & 2.258 & 0.457 & 0.602 & 3.515 \\[3pt]
    Stable-Codec & 16000 & 4 & 1.0 & 2.195 & 0.011 & 1.979 & 1.534 & 1.628 & 0.403 & 2.708 & 0.489 & 0.625 & 3.551 \\
    \vspace{0.05pt} \\[-0.05pt]
    \hline
    \vspace{0.05pt} \\[-0.05pt]
    X-codec-2.0 & 16000 & 1 & 0.8 & 1.380 & 0.018 & 2.056 & 1.591 & 0.685 & -6.592 & 0.562 & 0.651 & 0.640 & 3.730 \\
    \vspace{0.05pt} \\[-0.05pt]
    \hline
    \vspace{0.05pt} \\[-0.05pt]
    FlowDec & 48000 & 8 & 6.0 & 0.810 & 0.003 & 3.106 & 2.692 & 0.343 & 7.025 & 6.845 & 0.817 & 0.823 & 4.274 \\[3pt]
    FlowDec & 48000 & 10 & 7.5 & 0.762 & 0.003 & 3.336 & 2.984 & 0.313 & 8.070 & 7.667 & 0.841 & 0.856 & 4.306 \\
    \vspace{0.05pt} \\[-0.05pt]
    \hline
    \vspace{0.05pt} \\[-0.05pt]
    Baichuan & 16000 & 8 & 1.075 & 1.300 & 0.024 & 1.865 & 1.489 & 0.843 & -9.381 & 0.032 & 0.601 & 0.567 & 3.402 \\
    \vspace{0.05pt} \\[-0.05pt]
    \hline
  \end{tabular}%
  }
  \caption{Comparisons between different codecs in CodecBench datasets, where \textbf{nq} refers to the number of quantizers and Baichuan refers to Baichuan-Audio tokenizer.}
  \label{tab:codecbench_performance}
\end{table}

%% file: Selected_Metric.tex
\begin{table}[ht]
  \centering
  \footnotesize
  \setlength{\tabcolsep}{2.5pt}
  \renewcommand{\arraystretch}{0.5}
  \resizebox{\textwidth}{!}{%
  \begin{tabular}{l c *{12}{c}}
    \toprule
    & & \multicolumn{3}{c}{\textbf{Speech}} & \multicolumn{3}{c}{\textbf{Music}} & \multicolumn{3}{c}{\textbf{Sounds}} & \multicolumn{3}{c}{\textbf{General Audio}} \\
    \cmidrule(lr){3-5} \cmidrule(lr){6-8} \cmidrule(lr){9-11} \cmidrule(lr){12-14}
    \textbf{Model} & \textbf{kbps} & \textbf{Mel Loss$\downarrow$} & \textbf{PESQ-NB$\uparrow$} & \textbf{STOI$\uparrow$} & \textbf{Mel Loss$\downarrow$} & \textbf{PESQ-NB$\uparrow$} & \textbf{STOI$\uparrow$} & \textbf{Mel Loss$\downarrow$} & \textbf{PESQ-NB$\uparrow$} & \textbf{STOI$\uparrow$} & \textbf{Mel Loss$\downarrow$} & \textbf{PESQ-NB$\uparrow$} & \textbf{STOI$\uparrow$} \\
    \midrule
    \vspace{0.05pt} \\[-0.05pt]
    DAC-24-8 & 6.0 & 0.634 & 3.546 & 0.914 & 0.820 & 3.608 & 0.717 & 0.657 & 3.331 & 0.833 & 0.683 & 3.413 & 0.869 \\[3pt]
    DAC-24-32 & 24.0 & \textbf{0.296} & \textbf{4.415} & \textbf{0.988} & \textbf{0.451} & \textbf{4.355} & \textbf{0.902} & \textbf{0.311} & \textbf{4.372} & \textbf{0.956} & \textbf{0.406} & \textbf{4.232} & \textbf{0.961} \\[3pt]
    DAC-44-9 & 77.4 & 0.565 & 3.977 & 0.941 & 0.729 & 3.906 & 0.776 & 0.597 & 3.759 & 0.865 & 0.616 & 3.871 & 0.903 \\
    \vspace{0.05pt} \\[-0.05pt]
    \hline
    \vspace{0.05pt} \\[-0.05pt]
    MaskGCT Codec-8 & 4.0 & 0.874 & 2.899 & 0.886 & 1.441 & 2.288 & 0.567 & 1.053 & 0.013 & 0.752 & 0.898 & 2.703 & 0.811 \\[3pt]
    MaskGCT Codec-12 & 6.0 & 0.838 & 3.032 & 0.900 & 1.400 & 2.393 & 0.599 & 1.021 & 0.013 & 0.774 & 0.860 & 2.828 & 0.831 \\
    \vspace{0.05pt} \\[-0.05pt]
    \hline
    \vspace{0.05pt} \\[-0.05pt]
    BigCodec & 1.04 & 0.950 & 2.403 & 0.820 & 1.698 & 1.838 & 0.438 & 1.559 & 1.938 & 0.615 & 1.205 & 2.201 & 0.706 \\
    \vspace{0.05pt} \\[-0.05pt]
    \hline
    \vspace{0.05pt} \\[-0.05pt]
    Mimi-8 & 1.1 & 1.254 & 2.302 & 0.799 & 1.707 & 2.223 & 0.462 & 1.437 & 2.068 & 0.646 & 1.311 & 2.185 & 0.708 \\[3pt]
    Mimi-32 & 4.4 & 1.001 & 3.282 & 0.901 & 1.505 & 2.997 & 0.655 & 1.251 & 2.701 & 0.788 & 1.069 & 3.102 & 0.845 \\
    \vspace{0.05pt} \\[-0.05pt]
    \hline
    \vspace{0.05pt} \\[-0.05pt]
    Stable-Codec-1 & 0.4 & 1.807 & 1.977 & 0.728 & 2.579 & 1.651 & 0.322 & 2.332 & 1.642 & 0.520 & 2.007 & 1.783 & 0.588 \\[3pt]
    Stable-Codec-4 & 1.0 & 1.778 & 2.115 & 0.754 & 2.539 & 1.698 & 0.348 & 2.295 & 1.703 & 0.548 & 1.959 & 1.926 & 0.618 \\
    \vspace{0.05pt} \\[-0.05pt]
    \hline
    \vspace{0.05pt} \\[-0.05pt]
    X-codec-2.0 & 0.8 & 0.974 & 2.310 & 0.798 & 1.808 & 1.730 & 0.393 & 1.554 & 1.912 & 0.620 & 1.145 & 2.172 & 0.696 \\
    \vspace{0.05pt} \\[-0.05pt]
    \hline
    \vspace{0.05pt} \\[-0.05pt]
    FlowDec-8 & 6.0 & 0.723 & 3.229 & 0.897 & 0.877 & 3.940 & 0.682 & 0.847 & 2.799 & 0.812 & 0.877 & 2.816 & 0.785 \\[3pt]
    FlowDec-10 & 7.5 & 0.671 & 3.494 & 0.922 & 0.849 & 4.065 & 0.713 & 0.798 & 2.996 & 0.849 & 0.827 & 3.098 & 0.831 \\
    \vspace{0.05pt} \\[-0.05pt]
    \hline
    \vspace{0.05pt} \\[-0.05pt]
    Baichuan-audio & 1.075 & 1.106 & 1.896 & 0.726 & 1.444 & 1.980 & 0.397 & 1.436 & 1.782 & 0.520 & 1.095 & 1.943 & 0.652 \\
    \vspace{0.05pt} \\[-0.05pt]
    \hline
  \end{tabular}%
  }
  \caption{Comparisons between different codecs across four dataset categories: Speech, Music, Sound and General Audio.}
  \label{tab:selected_comprehensive}
\end{table}

%% file: Classification_Eval.tex
\begin{table}[ht]
  \centering
  \footnotesize
  \setlength{\tabcolsep}{2.2pt}
  \renewcommand{\arraystretch}{0.5}
  \resizebox{\textwidth}{!}{%
  \begin{tabular}{l c c c c c c c c c c}
    \noalign{\global\arrayrulewidth 1pt}\hline \vspace{0.05pt} \\[-0.05pt]\noalign{\global\arrayrulewidth 0.4pt}
    & & & \multicolumn{3}{c}{\textbf{Speech}} & \multicolumn{3}{c}{\textbf{Music}} & \multicolumn{2}{c}{\textbf{Sound}} \\
    \cmidrule(lr){4-6} \cmidrule(lr){7-9} \cmidrule(lr){10-11}
    {\textbf{Model}} & {\textbf{nq}} & {\textbf{Hidden size}}  & \textbf{RAVDESS} & \textbf{CREMA-D} & \textbf{MELD} & \textbf{Chord} & \textbf{GTZAN} & \textbf{Nsynth} & \textbf{ESC-50} & \textbf{VocalSound} \\
    \vspace{0.05pt} \\[-0.05pt]
    \hline
    \vspace{0.05pt} \\[-0.05pt]
    Whisper-Small & / & 768 & \textbf{0.690} & \textbf{0.608} & \textbf{0.527} & 0.575 & 0.687 & 0.592 & \textbf{0.567} & \textbf{0.904} \\
    \vspace{0.05pt} \\[-0.05pt]
    \hline
    \vspace{0.05pt} \\[-0.05pt]
    DAC 24k & 8 & 1024 & 0.413 & 0.377 & 0.471 & 0.575 & 0.520 & 0.565 & 0.257 & 0.426 \\[2pt]
    DAC 24k & 32 & 1024 & 0.440 & 0.359 & 0.423 & 0.579 & 0.556 & 0.642 & 0.270 & 0.430 \\
    \vspace{0.05pt} \\[-0.05pt]
    \hline
    \vspace{0.05pt} \\[-0.05pt]
    MaskGCT Codec & 8 & 256 & 0.495 & 0.405 & 0.485 & 0.584 & 0.565 & 0.615 & 0.332 & 0.486 \\[2pt]
    MaskGCT Codec & 10 & 256 & 0.511 & 0.419 & 0.487 & 0.585 & 0.589 & 0.635 & 0.359 & 0.504 \\
    \vspace{0.05pt} \\[-0.05pt]
    \hline
    \vspace{0.05pt} \\[-0.05pt]
    BigCodec & 1 & 1024 & 0.327 & 0.370 & 0.482 & 0.579 & 0.442 & 0.435 & 0.206 & 0.346 \\
    \vspace{0.05pt} \\[-0.05pt]
    \hline
    \vspace{0.05pt} \\[-0.05pt]
    Mimi & 8 & 512 & 0.621 & 0.512 & 0.487 & 0.588 & 0.721 & 0.687 & 0.547 & 0.796 \\[2pt]
    Mimi & 32 & 512 & 0.617 & 0.497 & 0.474 & \textbf{0.610} & \textbf{0.724} & \textbf{0.729} & 0.549 & 0.796 \\
    \vspace{0.05pt} \\[-0.05pt]
    \hline
    \vspace{0.05pt} \\[-0.05pt]
    X-codec-2.0 & 1 & 1024 & 0.332 & 0.335 & 0.484 & 0.574 & 0.430 & 0.425 & 0.230 & 0.336 \\
    \vspace{0.05pt} \\[-0.05pt]
    \hline
    \vspace{0.05pt} \\[-0.05pt]
    Baichuan & 8 & 1280 & 0.558 & 0.455 & 0.495 & 0.592 & 0.702 & 0.631 & 0.558 & 0.763 \\
    \vspace{0.05pt} \\[-0.05pt]
    \hline
  \end{tabular}%
  }
  \caption{Comparisons between different codecs on classification task. The indicator is the classification accuracy on the dataset. Baichuan refers to Baichuan-Audio tokenizer and Chord refers to the Musical Instrument Chord Classification dataset.}
  \label{tab:classification_performance}
\end{table}

%% file: 500_Limitaions.tex
Although CodecBench integrates numerous open-source datasets and incorporates our self-collected dataset to better align with real-world requirements, it still falls short in addressing certain extreme scenarios encountered in real-world applications. More tests of audio codecs under such conditions require further development. Additionally, current methods for evaluating semantic information are still not enough. The two embedding-based approaches employed are relatively simplistic and fail to capture the full spectrum of semantic information, which is inherently broad and multifaceted. In future work, we plan to expand the evaluation for semantic assessment to address these shortcomings more effectively.

%% file: 600_Conclusion.tex
This paper introduces CodecBench, a comprehensive evaluation framework for audio codecs, focusing on both acoustic and semantic dimensions. To meet the growing demands of advanced speech language models for high-quality audio codecs, we collected multiple datasets that better reflect real-world scenarios and incorporated self-collected data to enhance acoustic evaluation. For semantic evaluation, we employed two embedding-based methods inspired by prior work, which provide a more in-depth assessment compared to traditional benchmarks. Additionally, we conducted an extensive analysis of several recent codec models, showing the strengths and weaknesses of different architectures and approaches across various dimensions. Finally, we have released our code and plan to further refine CodecBench, fostering development within the community.

%% file: 900_Appendix.tex
\section{Datasets Details}
\label{appendix:datasets_details}
CodecBench has 18 open-source datasets and 1 self-collected dataset, including 6 speech datasets, 3 music datasets, 5 sound datasets and 5 general audio datasets. Licenses of open-source datasets re shown in Table~\ref{tab:codecbench_licenses}.

\input{Dataset_License}

\subsection{Speech}

\textbf{KeSpeech} KeSpeech~\citep{tang2021kespeech} is an open-source speech dataset comprising speech signals recorded by 27,237 speakers across 34 cities in China. The dataset includes standard Mandarin and its eight subdialects. It features multiple labels, including content transcription, speaker identity, and subdialect, supporting tasks such as speech recognition, speaker verification, subdialect identification, multi-task learning, and conditional learning.

\textbf{LibriSpeech} LibriSpeech~\citep{panayotov2015librispeech} is a highly utilized corpus of English speech data, comprising approximately 1000 hours of audio recordings. These recordings are characterized by a reading style, as they consist of utterances read from audiobooks.

\textbf{Libri2Mix} Libri2Mix~\citep{cosentino2020librimix} is a synthesized corpus that features mixtures of the speech of two speakers intertwined with background noise. The speech segments are sourced from LibriSpeech, the ambient noise is taken from the WHAM! dataset. We used the test clean set.

\textbf{MELD} MELD~\citep{poria2018meld} contains more than 1400 dialogues and 13000 utterances from the Friends TV series. Multiple speakers participated in the dialogues. Each utterance in a dialogue has been labeled by emotions.

\textbf{RAVDESS} RAVDESS~\citep{livingstone2018ryerson} contains 24 professional actors (12 female, 12 male), who voice two lexically matched statements in a neutral North American accent. Speech includes calm, happy, sad, angry, fearful, surprise, and disgust expressions, and song contains calm, happy, sad, angry, and fearful emotions. Each expression is produced at two levels of emotional intensity (normal, strong), with an additional neutral expression.

\textbf{CREMA-D} CREMA-D~\citep{cao2014crema} has 7,442 original clips from 91 actors (43 female and 48 male). Each clip is annotated with six distinct emotions. Professional actors, guided by experienced theatre directors, skillfully express a designated emotion while delivering specific sentences. We used the test clean set.

\subsection{Music} 

\textbf{Nsynth} Nsynth~\citep{engel2017neural} stands out as a large-scale and high-quality collection of musical notes, significantly exceeding similar public datasets in size.

\textbf{GTZAN} GTZAN~\citep{sturm2013gtzan} includes music samples categorized into 10 genres, each containing 100 audio files. All audio files within the dataset have a standardized length of 30 seconds. 

\textbf{Musical Instrument Chord Classification} Musical Instrument Chord Classification~\citep{kaggle2024musicalinstruments} includes 859 major audio and minor audio of piano and guitar.

\subsection{Sound}

\textbf{Laughterscape} Laughterscape~\citep{xin2023laughter} is a corpus of 11,413 laughter sounds from 584 Japanese speakers, collected from YouTube. We used a subset of 8170 entries from this dataset.

\textbf{VocalSound} VocalSound~\citep{gong2022vocalsound} is a free dataset consisting of 21,024 crowd-sourced recordings of laughter, sighs, coughs, throat clearing, sneezes, and sniffs from 3,365 unique subjects. We use a subset of 3591 entries from this dataset.

\textbf{ESC-50} ESC-50~\citep{piczak2015esc} encompasses 2000 environmental sounds categorized into 50 classes. The clips within this dataset are manually selected from public field recordings compiled by the Freesound.org project.

\textbf{CatDog} CatDog~\citep{kaggle2024CatDogs} dataset contains 164 recordings of cat sounds (1,323 seconds) and 113 recordings of dog sounds (598 seconds).

\textbf{Gunshot Triangulation} Gunshot Triangulation~\citep{cooper2020gunshots} collects the audio of seven distinct firearms—comprising four pistols and three rifles—each fired a minimum of three times. The shots were directed sequentially toward a target positioned 45 meters away from the shooter in an open field.

\subsection{General Audio}

\textbf{AudioSet} AudioSet~\citep{gemmeke2017audio} consists of segments of approximately 10 seconds each, of YouTube video labeled with over 500 audio events, featuring diverse environments and sound qualities.

\textbf{Air-Bench Chat} Air-Bench~\citep{yang2024air} encompasses two dimensions: foundation and chat benchmarks. The former consists of 19 tasks with approximately 19k single-choice questions. The latter one contains open-ended question-and-answer data. We use the chat benchmark's dataset.

\textbf{WavCaps Soundbible} WavCaps Soundbible~\citep{mei2024wavcaps} is a large-scale weakly-labelled audio captioning dataset, comprising 1232 audio clips with paired captions sourced from Soundbible.

\textbf{Clotho-AQA} Clotho-AQA~\citep{lipping2022clotho} is a dataset for Audio question answering consisting of 1991 audio files each between 15 to 30 seconds duration selected from the Clotho dataset.

\textbf{Self-collected Dataset} We collected more exaggerated and complex speaker scenes from the Bilibili, such as quarrels and speech with background music and vocal, which have higher requirements for audio codec performance. The dataset contains 400 entries.

\section{ASR Probing Task Details}
\label{appendix:asr_probing_details}
To enable effective alignment in the Automatic Speech Recognition (ASR) probing task, particularly for low-bitrate codecs, \textbf{we upsample the embeddings for models with a frame rate below 50 Hz to a minimum of 50 Hz using replication}. This upsampling is necessary because an insufficient input sequence length (\( T \)) relative to the target sequence length (\( U \)) can prevent the Connectionist Temporal Classification (CTC) loss from effectively aligning the input sequence (quantized features) with the target sequence (transcription characters). Specifically, CTC requires \( T \geq U \) to accommodate at least one time step per target label, and in the worst case, \( T \geq 2U + 1 \) to account for potential blank labels between each target label and at the sequence boundaries. Upsampling ensures that \( T \) is sufficiently large, particularly for low-frame-rate codes, to satisfy these constraints and enable effective alignment.

\section{Additional Experiment Results}
\label{appendix:Additional experiment results}

\input{Speech_Average}

\input{Music_Average}

\input{Sound_Average}

\input{General_Average}

%% file: Dataset_License.tex
\begin{table}[ht]
  \centering
  \footnotesize
  \setlength{\tabcolsep}{4pt}
  \begin{tabular}{l l}
    \toprule
    \textbf{Speech Dataset} & \textbf{License} \\
    \midrule
    KeSpeech & CC BY-NC-SA \\
    LibriSpeech & CC BY 4.0 \\
    Libri2Mix & MIT License \\
    MELD & GPL-3.0 License \\
    CREMA-D & Open Database License \\
    RAVDESS & CC BY-NC-SA 4.0 \\
    \midrule
    \textbf{Music Dataset} & \textbf{License} \\
    \midrule
    NSynth & CC BY 4.0 \\
    GTZAN & CC BY 4.0, Apache License v.2.0 \\
    Musical Instrument Chord Classification & CC0 1.0 \\
    \midrule
    \textbf{Sound Dataset} & \textbf{License} \\
    \midrule
    Laughterscape & Research and development purpose only (tentative) \\
    VocalSound & CC BY-SA 4.0 \\
    ESC-50 & CC BY-NC 3.0 \\
    CatDog & CC BY-SA 3.0 \\
    Gunshot Triangulation & CC0 1.0 \\
    \midrule
    \textbf{General Audio Dataset} & \textbf{License} \\
    \midrule
    AudioSet & CC BY-SA 4.0 \\
    Air-Bench Chat & Apache License Version 2.0 \\
    WavCaps Soundbible & CC BY 4.0 \\
    Clotho-AQA & MIT License \\
    \bottomrule
  \end{tabular}
  \caption{Licenses for open-source datasets used in CodecBench.}
  \label{tab:codecbench_licenses}
\end{table}

%% file: Speech_Average.tex
\begin{table}[ht]
  \centering
  \footnotesize
  \setlength{\tabcolsep}{3pt}
  \renewcommand{\arraystretch}{0.5}
  \resizebox{\textwidth}{!}{%
  \begin{tabular}{l c c c c c c c c c c c c c}
    \noalign{\global\arrayrulewidth 1pt}\hline\noalign{\global\arrayrulewidth 0.4pt}
    \vspace{0.05pt} \\[-0.05pt]
    \textbf{Model} & \textbf{Sample Rate} & \textbf{nq} & \textbf{kbps} & \textbf{Mel Loss$\downarrow$} & \textbf{MSE$\downarrow$} & \textbf{PESQ-NB$\uparrow$} & \textbf{PESQ-WB$\uparrow$} & \textbf{SC$\downarrow$} & \textbf{SDR$\uparrow$} & \textbf{SI-SDR$\uparrow$} & \textbf{SIM$\uparrow$} & \textbf{STOI$\uparrow$} & \textbf{ViSQOL$\uparrow$} \\
    \vspace{0.05pt} \\[-0.05pt]
    \hline
    \vspace{0.05pt} \\[-0.05pt]
    DAC & 24000 & 8 & 6.0 & 0.634 & 0.004 & 3.546 & 3.150 & 0.276 & 2.905 & 1.077 & 0.823 & 0.914 & 4.238 \\[3pt]
    DAC & 24000 & 32 & 24.0 & \textbf{0.296} & 0.004 & \textbf{4.415} & \textbf{4.438} & \textbf{0.115} & 3.604 & 1.110 & \textbf{0.967} & \textbf{0.988} & \textbf{4.450} \\[3pt]
    DAC & 44000 & 9 & 7.74 & 0.565 & \textbf{0.0003} & 3.977 & 3.673 & 0.210 & \textbf{11.186} & \textbf{10.536} & 0.893 & 0.941 & 4.414 \\
    \vspace{0.05pt} \\[-0.05pt]
    \hline
    \vspace{0.05pt} \\[-0.05pt]
    MaskGCT Codec & 24000 & 8 & 4.0 & 0.874 & 0.002 & 2.899 & 2.240 & 0.411 & 0.821 & 2.867 & 0.815 & 0.886 & 4.113 \\[3pt]
    MaskGCT Codec & 24000 & 12 & 6.0 & 0.838 & 0.002 & 3.032 & 2.371 & 0.394 & 1.428 & 3.193 & 0.848 & 0.900 & 4.150 \\
    \vspace{0.05pt} \\[-0.05pt]
    \hline
    \vspace{0.05pt} \\[-0.05pt]
    BigCodec & 16000 & 1 & 1.04 & 0.950 & 0.003 & 2.403 & 1.830 & 0.476 & -0.859 & 1.989 & 0.647 & 0.820 & 4.072 \\
    \vspace{0.05pt} \\[-0.05pt]
    \hline
    \vspace{0.05pt} \\[-0.05pt]
    Mimi & 24000 & 8 & 1.1 & 1.254 & 0.002 & 2.302 & 1.774 & 0.449 & 1.458 & 2.938 & 0.604 & 0.799 & 3.881 \\[3pt]
    Mimi & 24000 & 32 & 4.4 & 1.001 & 0.001 & 3.282 & 2.777 & 0.281 & 7.767 & 7.089 & 0.846 & 0.901 & 4.086 \\
    \vspace{0.05pt} \\[-0.05pt]
    \hline
    \vspace{0.05pt} \\[-0.05pt]
    Stable-Codec & 16000 & 1 & 0.4 & 1.807 & 0.005 & 1.977 & 1.528 & 2.436 & 0.775 & 2.601 & 0.370 & 0.728 & 3.681 \\[3pt]
    Stable-Codec & 16000 & 4 & 1.0 & 1.778 & 0.005 & 2.115 & 1.617 & 2.436 & 2.254 & 3.306 & 0.416 & 0.754 & 3.712 \\
    \vspace{0.05pt} \\[-0.05pt]
    \hline
    \vspace{0.05pt} \\[-0.05pt]
    X-codec-2.0 & 16000 & 1 & 0.8 & 0.974 & 0.003 & 2.310 & 1.761 & 0.519 & -2.389 & 1.205 & 0.673 & 0.798 & 4.018 \\
    \vspace{0.05pt} \\[-0.05pt]
    \hline
    \vspace{0.05pt} \\[-0.05pt]
    FlowDec & 48000 & 8 & 6.0 & 0.723 & 0.001 & 3.229 & 2.774 & 0.307 & 7.069 & 6.755 & 0.789 & 0.897 & 4.366 \\[3pt]
    FlowDec & 48000 & 10 & 7.5 & 0.671 & 0.001 & 3.494 & 3.133 & 0.272 & 8.228 & 7.637 & 0.821 & 0.922 & 4.398 \\
    \vspace{0.05pt} \\[-0.05pt]
    \hline
    \vspace{0.05pt} \\[-0.05pt]
    Baichuan & 16000 & 8 & 1.075 & 1.106 & 0.007 & 1.896 & 1.488 & 0.767 & -7.939 & 0.035 & 0.534 & 0.726 & 3.154 \\
    \vspace{0.05pt} \\[-0.05pt]
    \hline
  \end{tabular}%
  }
  \caption{Comparisons between different codecs in CodecBench datasets for Speech category, where Baichuan refers to Baichuan-Audio tokenizer.}
  \label{tab:codecbench_performance_speech}
\end{table}

%% file: Music_Average.tex
\begin{table}[ht]
  \centering
  \footnotesize
  \setlength{\tabcolsep}{3pt}
  \renewcommand{\arraystretch}{0.5}
  \resizebox{\textwidth}{!}{%
  \begin{tabular}{l c c c c c c c c c c c c c}
    \noalign{\global\arrayrulewidth 1pt}\hline\noalign{\global\arrayrulewidth 0.4pt}
    \vspace{0.05pt} \\[-0.05pt]
    \textbf{Model} & \textbf{Sample Rate} & \textbf{nq} & \textbf{kbps} & \textbf{Mel Loss$\downarrow$} & \textbf{MSE$\downarrow$} & \textbf{PESQ-NB$\uparrow$} & \textbf{PESQ-WB$\uparrow$} & \textbf{SC$\downarrow$} & \textbf{SDR$\uparrow$} & \textbf{SI-SDR$\uparrow$} & \textbf{SIM$\uparrow$} & \textbf{STOI$\uparrow$} & \textbf{ViSQOL$\uparrow$} \\
    \vspace{0.05pt} \\[-0.05pt]
    \hline
    \vspace{0.05pt} \\[-0.05pt]
    DAC & 24000 & 8 & 6.0 & 0.820 & 0.023 & 3.608 & 3.196 & 0.242 & 9.530 & 2.926 & 0.881 & 0.717 & 4.391 \\[3pt]
    DAC & 24000 & 32 & 24.0 & \textbf{0.451} & 0.023 & \textbf{4.355} & \textbf{4.316} & \textbf{0.098} & 14.382 & 3.146 & \textbf{0.972} & \textbf{0.902} & \textbf{4.552} \\[3pt]
    DAC & 44000 & 9 & 7.74 & 0.729 & \textbf{0.002} & 3.906 & 3.569 & 0.183 & \textbf{14.539} & \textbf{13.545} & 0.927 & 0.776 & 4.463 \\
    \vspace{0.05pt} \\[-0.05pt]
    \hline
    \vspace{0.05pt} \\[-0.05pt]
    MaskGCT Codec & 24000 & 8 & 4.0 & 1.441 & 0.028 & 2.288 & 1.750 & 0.719 & -6.353 & 0.952 & 0.760 & 0.567 & 4.255 \\[3pt]
    MaskGCT Codec & 24000 & 12 & 6.0 & 1.400 & 0.028 & 2.393 & 1.815 & 0.708 & -5.858 & 1.041 & 0.788 & 0.599 & 4.297 \\
    \vspace{0.05pt} \\[-0.05pt]
    \hline
    \vspace{0.05pt} \\[-0.05pt]
    BigCodec & 16000 & 1 & 1.04 & 1.698 & 0.031 & 1.838 & 1.456 & 0.798 & -9.143 & 0.516 & 0.583 & 0.438 & 4.072 \\
    \vspace{0.05pt} \\[-0.05pt]
    \hline
    \vspace{0.05pt} \\[-0.05pt]
    Mimi & 24000 & 8 & 1.1 & 1.707 & 0.014 & 2.223 & 1.721 & 0.504 & 2.432 & 3.619 & 0.734 & 0.462 & 3.998 \\[3pt]
    Mimi & 24000 & 32 & 4.4 & 1.505 & 0.008 & 2.997 & 2.521 & 0.339 & 8.164 & 7.487 & 0.846 & 0.655 & 4.225 \\
    \vspace{0.05pt} \\[-0.05pt]
    \hline
    \vspace{0.05pt} \\[-0.05pt]
    Stable-Codec & 16000 & 1 & 0.4 & 2.579 & 0.023 & 1.651 & 1.350 & 0.824 & -5.889 & 1.132 & 0.519 & 0.322 & 3.620 \\[3pt]
    Stable-Codec & 16000 & 4 & 1.0 & 2.539 & 0.022 & 1.698 & 1.381 & 0.795 & -4.325 & 1.404 & 0.539 & 0.348 & 3.660 \\
    \vspace{0.05pt} \\[-0.05pt]
    \hline
    \vspace{0.05pt} \\[-0.05pt]
    X-codec-2.0 & 16000 & 1 & 0.8 & 1.808 & 0.038 & 1.730 & 1.362 & 0.899 & -14.061 & 0.071 & 0.610 & 0.393 & 3.967 \\
    \vspace{0.05pt} \\[-0.05pt]
    \hline
    \vspace{0.05pt} \\[-0.05pt]
    FlowDec & 48000 & 8 & 6.0 & 0.877 & 0.004 & 3.940 & 3.696 & 0.226 & 13.262 & 12.083 & 0.866 & 0.682 & 4.457 \\[3pt]
    FlowDec & 48000 & 10 & 7.5 & 0.849 & 0.004 & 4.065 & 3.856 & 0.206 & 14.377 & 13.259 & 0.877 & 0.713 & 4.474 \\
    \vspace{0.05pt} \\[-0.05pt]
    \hline
    \vspace{0.05pt} \\[-0.05pt]
    Baichuan & 16000 & 8 & 1.075 & 1.444 & 0.045 & 1.980 & 1.518 & 0.923 & -15.599 & 0.022 & 0.665 & 0.397 & 3.837 \\
    \vspace{0.05pt} \\[-0.05pt]
    \hline
  \end{tabular}%
  }
  \caption{Comparisons between different codecs in CodecBench datasets for Music category, where Baichuan refers to Baichuan-Audio tokenizer.}
  \label{tab:codecbench_performance_music}
\end{table}

%% file: Sound_Average.tex
\begin{table}[ht]
  \centering
  \footnotesize
  \setlength{\tabcolsep}{3pt}
  \renewcommand{\arraystretch}{0.5}
  \resizebox{\textwidth}{!}{%
  \begin{tabular}{l c c c c c c c c c c c c c}
    \noalign{\global\arrayrulewidth 1pt}\hline\noalign{\global\arrayrulewidth 0.4pt}
    \vspace{0.05pt} \\[-0.05pt]
    \textbf{Model} & \textbf{Sample Rate} & \textbf{nq} & \textbf{kbps} & \textbf{Mel Loss$\downarrow$} & \textbf{MSE$\downarrow$} & \textbf{PESQ-NB$\uparrow$} & \textbf{PESQ-WB$\uparrow$} & \textbf{SC$\downarrow$} & \textbf{SDR$\uparrow$} & \textbf{SI-SDR$\uparrow$} & \textbf{SIM$\uparrow$} & \textbf{STOI$\uparrow$} & \textbf{ViSQOL$\uparrow$} \\
    \vspace{0.05pt} \\[-0.05pt]
    \hline
    \vspace{0.05pt} \\[-0.05pt]
    DAC & 24000 & 8 & 6.0 & 0.657 & 0.015 & 3.331 & 2.914 & 0.361 & 3.166 & 1.567 & 0.843 & 0.833 & 4.073 \\[3pt]
    DAC & 24000 & 32 & 24.0 & \textbf{0.311} & 0.015 & \textbf{4.372} & \textbf{4.323} & \textbf{0.175} & 4.814 & 1.743 & \textbf{0.952} & \textbf{0.956} & 4.267 \\[3pt]
    DAC & 44000 & 9 & 7.74 & 0.597 & \textbf{0.002} & 3.759 & 3.386 & 0.288 & \textbf{9.391} & \textbf{8.626} & 0.888 & 0.865 & \textbf{4.311} \\
    \vspace{0.05pt} \\[-0.05pt]
    \hline
    \vspace{0.05pt} \\[-0.05pt]
    MaskGCT Codec & 24000 & 8 & 4.0 & 1.053 & 0.013 & 2.484 & 1.917 & 0.620 & -5.619 & 0.619 & 0.788 & 0.752 & 3.950 \\[3pt]
    MaskGCT Codec & 24000 & 12 & 6.0 & 1.021 & 0.013 & 2.607 & 2.026 & 0.605 & -5.196 & 0.673 & 0.815 & 0.774 & 3.987 \\
    \vspace{0.05pt} \\[-0.05pt]
    \hline
    \vspace{0.05pt} \\[-0.05pt]
    BigCodec & 16000 & 1 & 1.04 & 1.559 & 0.015 & 1.938 & 1.530 & 0.728 & -8.436 & 0.346 & 0.612 & 0.615 & 3.467 \\
    \vspace{0.05pt} \\[-0.05pt]
    \hline
    \vspace{0.05pt} \\[-0.05pt]
    Mimi & 24000 & 8 & 1.1 & 1.437 & 0.008 & 2.068 & 1.611 & 0.569 & -1.161 & 1.806 & 0.713 & 0.646 & 3.693 \\[3pt]
    Mimi & 24000 & 32 & 4.4 & 1.251 & 0.005 & 2.701 & 2.170 & 0.414 & 3.701 & 3.950 & 0.822 & 0.788 & 3.872 \\
    \vspace{0.05pt} \\[-0.05pt]
    \hline
    \vspace{0.05pt} \\[-0.05pt]
    Stable-Codec & 16000 & 1 & 0.4 & 2.332 & 0.011 & 1.642 & 1.338 & 1.224 & -4.189 & 1.082 & 0.499 & 0.520 & 3.195 \\[3pt]
    Stable-Codec & 16000 & 4 & 1.0 & 2.295 & 0.011 & 1.703 & 1.374 & 1.215 & -3.212 & 1.307 & 0.516 & 0.548 & 3.211 \\
    \vspace{0.05pt} \\[-0.05pt]
    \hline
    \vspace{0.05pt} \\[-0.05pt]
    X-codec-2.0 & 16000 & 1 & 0.8 & 1.554 & 0.017 & 1.912 & 1.512 & 0.775 & -9.505 & 0.105 & 0.646 & 0.620 & 3.422 \\
    \vspace{0.05pt} \\[-0.05pt]
    \hline
    \vspace{0.05pt} \\[-0.05pt]
    FlowDec & 48000 & 8 & 6.0 & 0.847 & 0.004 & 2.799 & 2.388 & 0.414 & 4.728 & 5.125 & 0.826 & 0.812 & 4.217 \\[3pt]
    FlowDec & 48000 & 10 & 7.5 & 0.798 & 0.004 & 2.996 & 2.632 & 0.386 & 5.551 & 5.701 & 0.842 & 0.849 & 4.255 \\
    \vspace{0.05pt} \\[-0.05pt]
    \hline
    \vspace{0.05pt} \\[-0.05pt]
    Baichuan & 16000 & 8 & 1.075 & 1.436 & 0.021 & 1.782 & 1.477 & 0.911 & -11.273 & 0.033 & 0.654 & 0.520 & 3.371 \\
    \vspace{0.05pt} \\[-0.05pt]
    \hline
  \end{tabular}%
  }
  \caption{Comparisons between different codecs in CodecBench datasets for Sound category, where Baichuan refers to Baichuan-Audio tokenizer.}
  \label{tab:codecbench_performance_sound}
\end{table}

%% file: General_Average.tex
\begin{table}[ht]
  \centering
  \footnotesize
  \setlength{\tabcolsep}{3pt}
  \renewcommand{\arraystretch}{0.5}
  \resizebox{\textwidth}{!}{%
  \begin{tabular}{l c c c c c c c c c c c c c}
    \noalign{\global\arrayrulewidth 1pt}\hline\noalign{\global\arrayrulewidth 0.4pt}
    \vspace{0.05pt} \\[-0.05pt]
    \textbf{Model} & \textbf{Sample Rate} & \textbf{nq} & \textbf{kbps} & \textbf{Mel Loss$\downarrow$} & \textbf{MSE$\downarrow$} & \textbf{PESQ-NB$\uparrow$} & \textbf{PESQ-WB$\uparrow$} & \textbf{SC$\downarrow$} & \textbf{SDR$\uparrow$} & \textbf{SI-SDR$\uparrow$} & \textbf{SIM$\uparrow$} & \textbf{STOI$\uparrow$} & \textbf{ViSQOL$\uparrow$} \\
    \vspace{0.05pt} \\[-0.05pt]
    \hline
    \vspace{0.05pt} \\[-0.05pt]
    DAC & 24000 & 8 & 6.0 & 0.683 & 0.018 & 3.413 & 2.991 & 0.293 & 2.170 & 1.540 & 0.886 & 0.869 & 4.136 \\[3pt]
    DAC & 24000 & 32 & 24.0 & \textbf{0.406} & 0.014 & \textbf{4.232} & \textbf{4.112} & \textbf{0.156} & 5.582 & 4.616 & \textbf{0.966} & \textbf{0.961} & \textbf{4.395} \\[3pt]
    DAC & 44000 & 9 & 7.74 & 0.616 & \textbf{0.002} & 3.871 & 3.517 & 0.228 & \textbf{10.335} & \textbf{10.085} & 0.932 & 0.903 & 4.366 \\
    \vspace{0.05pt} \\[-0.05pt]
    \hline
    \vspace{0.05pt} \\[-0.05pt]
    MaskGCT Codec & 24000 & 8 & 4.0 & 0.898 & 0.017 & 2.703 & 2.121 & 0.483 & -3.043 & 2.144 & 0.869 & 0.811 & 4.034 \\[3pt]
    MaskGCT Codec & 24000 & 12 & 6.0 & 0.860 & 0.017 & 2.828 & 2.239 & 0.467 & -2.654 & 2.358 & 0.888 & 0.831 & 4.076 \\
    \vspace{0.05pt} \\[-0.05pt]
    \hline
    \vspace{0.05pt} \\[-0.05pt]
    BigCodec & 16000 & 1 & 1.04 & 1.205 & 0.020 & 2.201 & 1.734 & 0.554 & -5.792 & 1.468 & 0.725 & 0.706 & 3.404 \\
    \vspace{0.05pt} \\[-0.05pt]
    \hline
    \vspace{0.05pt} \\[-0.05pt]
    Mimi & 24000 & 8 & 1.1 & 1.311 & 0.010 & 2.185 & 1.703 & 0.472 & 0.333 & 2.900 & 0.753 & 0.708 & 3.646 \\[3pt]
    Mimi & 24000 & 32 & 4.4 & 1.069 & 0.006 & 3.102 & 2.621 & 0.321 & 6.014 & 6.167 & 0.903 & 0.845 & 3.852 \\
    \vspace{0.05pt} \\[-0.05pt]
    \hline
    \vspace{0.05pt} \\[-0.05pt]
    Stable-Codec & 16000 & 1 & 0.4 & 2.007 & 0.016 & 1.783 & 1.390 & 0.990 & -3.585 & 1.907 & 0.454 & 0.588 & 3.051 \\[3pt]
    Stable-Codec & 16000 & 4 & 1.0 & 1.959 & 0.015 & 1.926 & 1.483 & 0.964 & -2.041 & 2.505 & 0.498 & 0.618 & 3.084 \\
    \vspace{0.05pt} \\[-0.05pt]
    \hline
    \vspace{0.05pt} \\[-0.05pt]
    X-codec-2.0 & 16000 & 1 & 0.8 & 1.145 & 0.023 & 2.172 & 1.699 & 0.598 & -8.500 & 0.918 & 0.769 & 0.696 & 3.365 \\
    \vspace{0.05pt} \\[-0.05pt]
    \hline
    \vspace{0.05pt} \\[-0.05pt]
    FlowDec & 48000 & 8 & 6.0 & 0.877 & 0.008 & 2.816 & 2.368 & 0.378 & 5.481 & 5.955 & 0.826 & 0.785 & 4.063 \\[3pt]
    FlowDec & 48000 & 10 & 7.5 & 0.827 & 0.007 & 3.098 & 2.691 & 0.347 & 6.604 & 6.746 & 0.853 & 0.831 & 4.095 \\
    \vspace{0.05pt} \\[-0.05pt]
    \hline
    \vspace{0.05pt} \\[-0.05pt]
    Baichuan & 16000 & 8 & 1.075 & 1.095 & 0.032 & 1.943 & 1.513 & 0.750 & -13.301 & 0.021 & 0.740 & 0.652 & 3.501 \\
    \vspace{0.05pt} \\[-0.05pt]
    \hline
  \end{tabular}%
  }
  \caption{Comparisons between different codecs in CodecBench datasets for General Audio category, where Baichuan refers to Baichuan-Audio tokenizer.}
  \label{tab:codecbench_performance_general}
\end{table}